\documentclass{eas}
\newcommand{\al}{Al$_2$O$_3$}
\newcommand{\sif}{Mg$_2$SiO$_4$}
\newcommand{\sie}{MgSiO$_3$}
\newcommand{\spi}{Mg$_2$Al$_2$O$_4$}
\newcommand{\bet}{$\alpha$ Ori}
\newcommand{\Ms}{M$_{\odot}$}
\newcommand{\Rs}{R$_{\odot}$}

\usepackage{graphicx}
\begin{document}

\title{The chemistry of dust formation in red supergiants} 
\author{Isabelle Cherchneff}\address{Physik Departement, Universit{\"a}t Basel, Klingelbergstrasse 82, CH-4056 Basel, Switzerland}
%
%
\begin{abstract}
Massive stars in their late stages of evolution as Red Supergiants experience mass loss. The resulting winds show various degrees of dynamical and chemical complexity and produce molecules and dust grains. This review summarises our knowledge of the molecular and dust components of the wind of Red Supergiants, including VY CMa and Betelgeuse. We discuss the synthesis of dust as a non equilibrium process in stellar winds, and present the current knowledge of the chemistry involved in the formation of oxygen-rich dust such as silicates and metal oxides. 
\end{abstract}
\runningtitle{Dust Formation in Red Supergiants}
\maketitle
\section{Introduction}
The discovery of large masses of dust in high-redshift quasars and Damped-Lyman-Alpha Systems two decades ago (\cite{pei91}, \cite{bert03}) rekindled the study of cosmic dust formation in evolved circumstellar environments (ECEs). While evolved objects on the Asymptotic Giant Branch (AGB) are by far the dominant dust makers in local galaxies and the Milky Way, other candidates contributing to the dust budget in the early universe include massive stars and their ultimate stage of evolution, their explosion as supernovae (SNe). The most abundant dust-forming SNe are of Type II-P and have Red Supergiant stars (hereafter RSGs) as progenitors. The interaction of the SN explosion blast wave with the circumstellar wind of the RSG progenitor indirectly probes the nature of the RSG wind in term of chemical composition, dust content, and mass loss rate. Understanding the way massive stars form dust thus becomes a crucial issue to the assessment of the galactic dust budget in the local and far Universe.  

RSGs correspond to the late stage of evolution of moderately massive stars ($10 \leq$ M$_{\star} < 30$ \Ms). Theses stars show a wide spread in their mass loss rates, with values ranging from $\sim 1 \times 10^{-6}$ \Ms~yr$^{-1}$~for \bet~(\cite{mau11}) to  $\sim 3 \times 10^{-4}$ \Ms~yr$^{-1}$~for VY CMa (\cite{hum05}). The mechanisms triggering mass loss are not well understood, and unlike the case of AGB stars, dust certainly plays a minor role to the overall mass loss process. RSG winds are probably accelerated in the stellar outer atmosphere by the combined effect of convection, rotation, and chromospheric activity (\cite{uit98}; \cite{joss07}; \cite{ohn11}), resulting in an extremely heterogeneous nebula where small to large convective cells, shocks, plumes and knots are observed (Kervella {\em et al.\/} 2009, 2011; \cite{chia10}). Understanding the synthesis of solid grains in RSGs thus represents an interesting challenge because of the dynamics and harsh conditions encountered in the stellar outer atmosphere and wind. 

\section{Molecules and dust in red supergiants}
A wealth of observational data exists on RSG winds in the (near-) infrared (IR). Molecules have been detected in the wind of several objects, including VY CMa,  $\mu$ Cep, and \bet.  The water molecule, H$_2$O, was first detected in $\mu$ Cep and \bet~by \cite{dan65}. However, Tsuji (2000, 2006) first recognised that the observed near-IR absorption bands originated from a molecular layer called the MOLsphere, consisting of water, and to a minor extent, the cyano radical, CN. The MOLsphere is found to be located above but close to the stellar photosphere, at $\sim$ 2 \Rs. For the specific case of \bet, a thin MOLsphere located at 1.3 \Rs~made of silicon monoxide, SiO, water, and alumina (\al) grains is necessary to reproduce the optical depth excess  in the mid-IR (\cite{per07}). 

Major advances on the composition of RSG winds have been achieved with the submm Herschel space telescope. In VY CMa, \cite{roy10} showed that the wind harboured a very rich, oxygen chemistry dominated by H$_2$O, where O-bearing species coexisted with carbon-bearing species such as HCN and CS. The presence of both oxygen- and carbon-bearing molecules may trace the breaking of CO molecules by shocks, as in the case of AGB stars (Cherchneff 2006, 2012), and/or the existence of a strong UV field. The great molecular richness of VY CMa was already emphasised by Ziurys {\it et al.} (2007) with the detection of many molecular rotational transitions tracing various components of the wind. Small metal oxide species including AlO, TiO, and TiO$_2$ have also been detected in the gas-phase close to the star (\cite{ten09}; Kami{\'n}ski {\em et al.\/} 2012, 2013), revealing a pool of gas-phase species involved in the formation of dust such as alumina and titanates. A similarly rich molecular outflow is found for the RSG NML Cygni, but the MOLsphere of \bet~shows a more limited molecular sample, including CO, H$_2$O and OH (\cite{tes12}). No SiO lines were detected in the Herschel spectra of \bet~reported by Teyssier {\em et al.\/} (2012), in contrast with the modelling of spectro-interferometric data by Perrin {\em et al.\/} (2007), for which a large SiO column density in the MOLsphere is required. SiO being a key species in the formation pathway to silicates, some transitions of this species will probably be detected in the wind of \bet\ as more Herschel data become available. The molecular species so far observed in a few RSGs are summarised in Table 1. 

\begin{table}
\caption{Observed molecules in the inner and intermediate wind of the RSG stars \bet, VY CMa, and NML Cyg. Data are taken from Ziurys et al. (2007), Tenenbaum \& Ziurys (2009), Teyssier et al. (2012), and Kami{\'n}ski et al. (2013).  }             
\label{tab4}     
\centering                          
\begin{tabular}{lccc }
\\       
\hline\hline                
 Species & \bet& NML Cyg  &VY CMa \\
 \hline
CO  &  X&X&X \\
H$_2$O  &  X&X&X \\   
OH  &  X&X&X \\ 
SiO &  &X&X \\
HCN &  & X&X \\
H$_2$S & & X & X \\
CS & &  & X\\
SiS & & & X\\
CN &  &  &X \\
PN & &  & X\\
AlO & & &X \\
TiO & & &X \\
TiO$_2$ & & &X \\
NaCl & &  & X\\
HCO$^+$ & & &X\\

\hline
\end{tabular}
 
\end{table}

This plethora of detected molecules indicates that RSG winds must be conducive to dust formation. However, the dust formation radius derived from data modelling in RSGs spans a wide value range. \cite{dan94} derived a large inner radius ($\sim 40$ \Rs) for the dust shell in \bet~from interferometric data, while the dust shell for VY CMa was found close to the star ($\sim 5$ \Rs). Such a location is encountered in AGB stars, where dust forms in the shocked gas layers above the stellar photosphere (\cite{cher06}). The dust composition in RSGs is still a matter of debate. ISO-SWS mid-IR spectra show bands at 9 $\mu$m and 18 $\mu$m, indicative of the presence of amorphous silicates, namely olivine. In the case of \bet, an extra source of mid-IR opacity is required to explain the observed mid-IR excess, and the best fit to the data is achieved when alumina (\al)~dust is added to the composition of the MOLsphere (\cite{ver06}; \cite{per07}). For a large RSG sample, \cite{spe00} and Verhoelst {\em et al.\/} (2006) show that the mid-IR excess can be fitted with a combination of O-rich dust, including olivine, Mg$_{0.8}$Fe$_{1.2}$SiO$_4$, melilite, Ca$_2$Al$_2$O$_7$, and \al. However, this choice of condensates is guided by condensation sequences derived from thermodynamic equilibrium (TE) and may then be questionable, as discussed in \S \ref{note}. 

\section{Dust formation as a non-equilibrium process in stellar outflows}
\label{note} 
There exists to date no satisfactory model of dust formation in ECEs. The great difficulty resides in the characterisation of the different steps involved in the formation of solid grains out of the gas phase. In this regard, modelling the production of cosmic carbon dust took advantage of the numerous information provided by the field of combustion chemistry on the formation of soot (\cite{frek89}; \cite{cher92}). As to silicate dust, notwithstanding the studies purveyed by the field of material science on ceramic and silicon film growth, the steps leading to the synthesis of silica and silicates are still partly unknown.  

In this context and for decades, studies on the synthesis of cosmic dust in various types of ECEs (e.g., AGB stars, RSGs, supernovae) have been based on the assumption of TE to predict "condensation sequences" (\cite{gros74}; \cite{tie90}; \cite{tie98}; \cite{lod06}), and the use of Classical Nucleation Theory (CNT) to derive grain size distributions for specific dust chemical compositions (\cite{koz89}; \cite{gail99}; \cite{tod01}; \cite{bian07}). For O-rich environments, the condensation sequence derived by Tielens (1990) and Tielens {\em et al.\/} (1998) is based on a description of the wind assuming TE from the photosphere to large radii, an assumption that does not hold in general in ECEs, owing to the complex dynamics of the outflows that include shocks, chromospheric activity, UV fields, etc. For example, alumina small clusters form after forsterite clusters in the O-rich zones of supernovae ejecta because of the radioactivity-induced chemistry specific to these environments (\cite{sar13}). This result is in direct contrast with the condensation sequences derived for O-rich environments (e.g., \cite{tie90}), and shows that the use of TE predictions can be misleading in the interpretation of complex media such as RSG winds. As to CNT, the description of the dust nucleation phase assumes steady state, optical properties and concepts (i.e., sticking coefficient, surface tension) appropriate for the bulk material. These assumptions are clearly not satisfactory to describe the formation of small molecular clusters entering the nucleation phase of dust synthesis for the conditions pertaining to ECEs. \cite{don85} were the first to question the use of CNT for the description of dust formation in space decades ago. 

More generally, dust grains form out of the gas phase following a two-step process that involves a nucleation and condensation phase (\cite{cher10}; \cite{cher11}; \cite{sar13}). The nucleation phase consists of the formation of molecules and molecular clusters out of the gas phase and is controlled by chemical kinetics and the ability of gas-phase species (i.e, atoms, molecules, ions) to react with one another to form larger chemical entities. Chemical reactions, their rate and occurrence, therefore control the growth of clusters from the gas phase. The condensation phase produces dust grains through the coalescence, surface growth, and coagulation of large molecular clusters. This phase is usually described by various formalisms (moments, bin, or Monte-Carlo methods) based on collision theory. Both phases are observed and studied in the laboratory with many experimental devices and techniques producing dust such as flame aerosol technologies and vaporisation experiments. 
 
\section{Nucleation of oxygen-rich dust}    
The formation of cosmic dust from a gaseous medium requires high gas densities and temperatures and sufficient time for the nucleation of small clusters to proceed, followed by their coalescence, coagulation, and growth to form dust grains. Therefore, the dense, hot zones above the photosphere of evolved stars such as AGBs or RSGs are conducive to the production of dust grains. Because the photosphere of RSGs is rich in oxygen, silicon, magnesium, and aluminium owing to the nucleosynthesis in massive stellar cores, the condensates should include calcium-magnesium-iron silicates (members of both the olivine and pyroxene family) and metal oxides (e.g., alumina \al~ and/or spinel \spi). Despite the wealth of studies dedicated to the formation of silicates and ceramics in the laboratory, the identification of molecular precursors to dust formation is still a difficult task. For silicates, alumina, or spinel, no gas-phase monomers exist from which solids grow by direct coalescence or coagulation. Flame aerosol or shock-induced combustion experiments have indeed highlighted the important role of heterogeneous seed clusters in the formation of these solid compounds (\cite{mcmil96}; \cite{prat98}; \cite{cat03}). 

\subsection{Silicates}  
\begin{figure}
\label{fig1}
\includegraphics[scale=0.38]{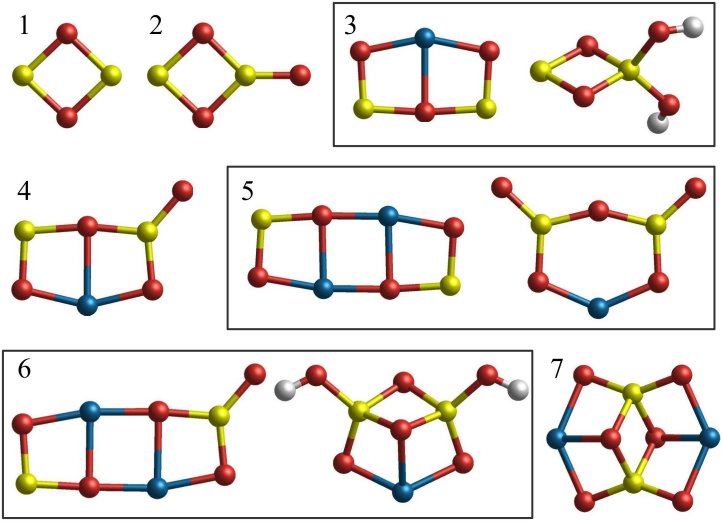}
\caption{The structures of a few clusters involved in the synthesis of forsterite dimers taken from \cite{gou12} (red: oxygen, yellow: silicon, blue: magnesium, grey: hydrogen). Cluster 7 corresponds to the entastite dimer (\sie)$_2$. }
\end{figure}

In an astrophysical context, silicate formation has been investigated in the laboratory by several groups, including Rietmeijer {\em et al.\/} (1999, 2002), \cite{kai03}, \cite{kam05}, and \cite{kim08}. In all experiments, the vaporisation of solid magnesium and silicon oxides led to the formation of crystalline Mg-bearing silicate like forsterite, \sif, Mg-bearing amorphous silicates, MgO grains (periclase and amorphous), SiO$_2$ grains and pure Si grains. Several studies emphasise a two-step process in which periclase and silica form in the first place, and further coalesce to give forsterite or amorphous silicates, e.g., \cite{kim08}. Under circumstellar conditions, \cite{kol97} showed that MgO grains were not the preferred intermediates in the dust nucleation process, a point corroborated by recent chemical models of supernova ejecta (\cite{sar13}). In the specific context of H-rich ECEs (e.g., AGB and RSG winds), the large H$_2$ content ensures that reaction of H$_2$ with MgO, a highly exothermic process, will destroy the newly formed small oxide, hampering its formation in the gas (\cite{pla13}). 

Recently, \cite{gou12} proposed a new nucleation scheme leading to small silicate clusters with the stochiometric composition of enstatite and forsterite. The scheme is based on the more energetically favourable paths to silicate growth. The growth mechanism consists of the initial formation of the SiO dimer, Si$_2$O$_2$, from a SiO-rich gas, followed by a two-step sequence involving 1) the addition of one oxygen atom to SiO dimers and subsequent silicate intermediates through reactions with O-bearing species available in the gas phase; and 2) the addition of one magnesium atom to the silicate cluster lattice. All chemical reactions characterising these addition steps are down-hill processes (exothermic), unlike the formation of the SiO dimer. Therefore, the closure of the Si$_2$O$_2$ ring represents the bottleneck to the synthesis of small silicate clusters. Some intermediate clusters involved in the formation of enstatite and forsterite dimers are illustrated in Figure 1.  

\subsection{Alumina} 
\begin{figure}
\label{fig2}
\includegraphics[scale=0.35]{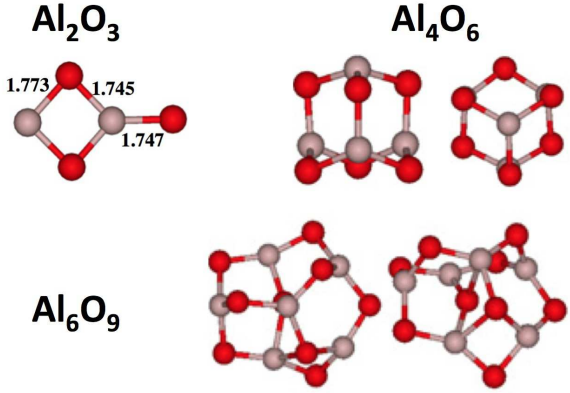}
\caption{Examples of the most stable structures of small \al~clusters adapted from Li \& Cheng (2012) (red: oxygen, grey: aluminium). }
\end{figure}

The structure of small alumina clusters has been extensively studied theoretically (e.g., \cite{archi99}; \cite{li12}), and some structures of small \al~polymer clusters are illustrated in Figure 2. Although there exists no well-identified chemical pathways to their formation, a possible route can be conjectured based on cluster structures. Because of the structural similarities between (SiO)$_2$ and (AlO)$_2$, the formation of small alumina clusters may first involve the closure of the (AlO)$_2$ ring, and the later addition of oxygen to lead to Al$_2$O$_3$. Further growth to \al~ dimers and trimers may occur with further addition of Al via reaction with Al, AlO, and AlOH, while oxygen addition may proceed from reaction with H$_2$O, OH, and/or O$_2$. Such a scheme awaits confirmation by a thourough assessment of the chemical pathways and rates involved in each growth step.   

\section{Silicate clusters in O-rich AGB winds}

\begin{figure}
\label{fig3}
\includegraphics[scale=0.42]{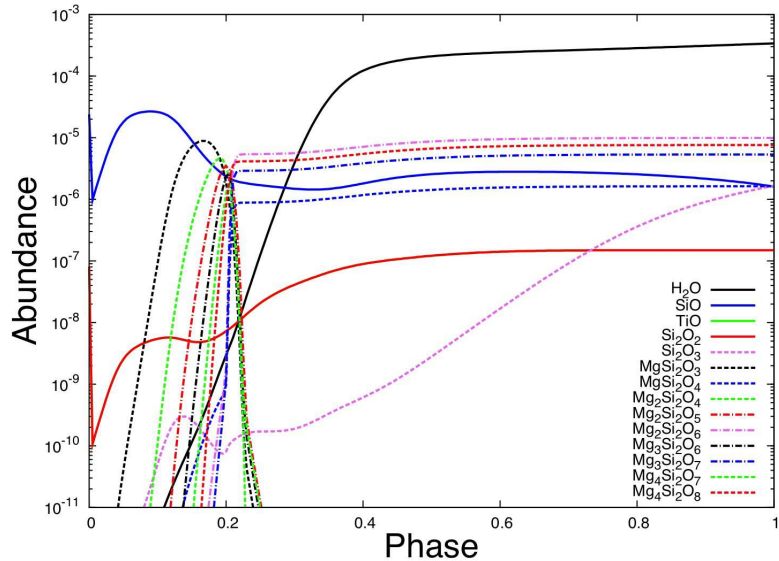}
\caption{Abundances of silicate clusters with respect to H$_2$ as a function of pulsation phase in the shocked inner wind of the Mira star IK Tau (Gobrecht \& Cherchneff, in preparation).   }
\end{figure}

The formation of silicate dust in RSGs chemically resembles that occurring in O-rich AGB stars. Both environments are H-rich and comprise silicon monoxide, SiO, atomic magnesium and iron, and water. To test the silicate chemical growth proposed by \cite{gou12}, the inner wind of the Mira star IK Tau is modelled following \cite{dua99} and Cherchneff (2006). The model includes a chemical kinetic description of the various gas-phase processes coupled to the nucleation routes to forsterite dimers, and considers the impact of shocks on the outer atmosphere of IK Tau. This approach was successful in reproducing the chemical composition of the inner wind of carbon- and O-rich AGBs (\cite{cher06, cher12}). The abundances of small silicate clusters with respect to H$_2$ as a function of pulsation phase in the post-shock gas above the photosphere are illustrated in Figure 3. One striking aspect is the rapid conversion of SiO to SiO dimers and the growth of clusters leading to the formation of forsterite and enstatite dimers when the post-shock gas temperature drops below $\sim$ 2500 K. This formation does not equally deplete all clusters, and small clusters such as SiO dimers and Mg$_3$Si$_2$O$_7$ may survive the nucleation step. These clusters may either be expelled as free-flying species in the wind or eventually coalesce at lower temperatures to form amorphous silicates. This preliminary study indicates that silicate dust clusters seem to easily form in the inner wind of Mira stars, pointing to the efficient formation of silicate dust in these environments.                                                                                                                                                                                                                                   
 
\section{Conclusions}
The gas conditions in RSG winds above the photosphere are hasher than those found in AGB outflows. As revealed by optical and UV observations of \bet, the chromospheric activity generates a hot plasma that coexists with a more abundant, cooler gas lifted above the photosphere by convective cells (\cite{lim98}). It is then likely that the formation of dust in RSGs occurs in the dense, cool, clumpy component of the wind. A puzzling case is that of \bet, where \al~dust is required in the MOLsphere close to the star to account for the mid-IR opacity excess, whilst silicate grains are observed at large radii above the stellar surface. Two scenarios have been proposed. The first scenario considers a two-shell configuration where dust destruction by chrosmospheric activity is effective after the formation of \al~until large radii and cooler temperatures are reached where anew silicate dust forms (\cite{ver09}). A second scenario considers the steady formation of \al~close to the photosphere. \al~being transparent, it would become detectable at radii where silicates form and coat the \al~grains, resulting in a silicate dust shell at large radii (\cite{ona89}). Both scenarios overlook the problem of the low gas density encountered at large radii which will hamper efficient silicate cluster coalescence and coagulation at these large distances. 

In some RSGs like VY CMa, the dust formation process may resemble that operating in AGB winds, where molecules and dust clusters are efficiently produced close to the star. For these objects, dust may contribute to the triggering of the mass loss close to the stellar surface. In the case of \bet, the low molecular content and poor chemical complexity observed close to the stellar photosphere (see Table~1) may indicate a less efficient dust production process. Dust there may not be a key player in the wind acceleration mechanism. More generally, dust formation is not an equilibrium process and is highly dependent of the physics and chemistry of the locus where grains form. Because the dust formation zone in RSGs is clumpy and has a complex dynamics, it is difficult to derive plausible scenarios for the formation of dust in such environments without a proper physico-chemical model of RSG winds.  


\end{document}